\documentclass[conference]{IEEEtran}
\IEEEoverridecommandlockouts
\usepackage{cite}
\usepackage{amsmath,amssymb,amsfonts}
\usepackage{algorithmic}
\usepackage{graphicx}
\usepackage{tikz}
\usepackage[left=2.55cm,right=2.55cm,top=1.9cm]{geometry}
\usetikzlibrary{positioning,fit,arrows.meta,backgrounds}

\tikzset{
    module/.style={%
        draw, rounded corners,
        minimum width=#1,
        minimum height=7mm,
        font=\sffamily
        },
    module/.default=2cm,
    >=LaTeX
}

\usepackage{textcomp}
\usepackage{xcolor}
\usepackage{hyperref}
\def\BibTeX{{\rm B\kern-.05em{\sc i\kern-.025em b}\kern-.08em
    T\kern-.1667em\lower.7ex\hbox{E}\kern-.125emX}}
\begin{document}
\newcommand{\franco}[1]{{\color{red}[Franco: #1]}}

\newcommand{\sofie}[1]{{\color{magenta}[Sofie: #1]}}
\newcommand{\robbert}[1]{{\color{green}[Robbert: #1]}}
\newcommand{\vida}[1]{{\color{blue}[vida: #1]}}
\newcommand{\rizqi}[1]{{\color{orange}[Rizqi: #1]}}
\newcommand{\ac}[1]{{\color{cyan}[Achiel: #1]}}
\newcommand{\cel}[1]{{\color{teal}[Cel: #1]}}

\newcommand{\hq}[1]{{\color{olive}[Haoqiu: #1]}}
\newcommand{\dieter}[1]{{\color{violet}[Dieter: #1]}}

\title{Building a real-time physical layer labeled data logging facility for 6G research

\thanks{This work is part of the 6G-BRICKS project, which has received funding from the Smart Networks and Services Joint Undertaking (SNS JU) under the European Union’s Horizon Europe research and innovation programme under Grant Agreement No 101096954.}
}

\author{
    \IEEEauthorblockN{
        Franco Minucci\IEEEauthorrefmark{1}, 
        Raquel Marina Noguera Oishi\IEEEauthorrefmark{1}, 
        Haoqiu Xiong\IEEEauthorrefmark{1}, 
        Dieter Verbruggen\IEEEauthorrefmark{1},
        Cel Thys\IEEEauthorrefmark{1},\\
        Rizqi Hersyandika\IEEEauthorrefmark{1},
        Robbert Beerten\IEEEauthorrefmark{1}, 
        Achiel Colpaert\IEEEauthorrefmark{2}\IEEEauthorrefmark{1}, 
        Vida Ranjbar\IEEEauthorrefmark{1}, 
        Sofie Pollin\IEEEauthorrefmark{2}\IEEEauthorrefmark{1}
        }

\IEEEauthorblockA{
    \textit{
        \IEEEauthorrefmark{1} Departement of Electrical Engineering (KU Leuven) 
        }\\
    \textit{
        \IEEEauthorrefmark{2}imec, Kapeldreef 75, 3001 Leuven, Belgium
        } \\
    }
}

\maketitle

\begin{abstract}
This work describes the architecture and vision of designing and implementing a new test infrastructure for 6G physical layer research at KU Leuven. 
The Testbed is designed for physical layer research and experimentation following several emerging trends, such as cell-free networking, integrated communication, sensing, open disaggregated Radio Access Networks, AI-Native design, and multiband operation.
The software is almost entirely based on free and open-source software, making contributing and reusing any component easy.
The open Testbed is designed to provide real-time and labeled data on all parts of the physical layer, from raw IQ data to synchronization statistics, channel state information, or symbol/bit/packet error rates. Real-time labeled datasets can be collected by synchronizing the physical layer data logging with a positioning and motion capture system.
One of the main goals of the design is to make it open and accessible to external users remotely. Most tests and data captures can easily be automated, and experiment code can be remotely deployed using standard containers (e.g., Docker or Podman).
Finally, the paper describes how the Testbed can be used for our research on joint communication and sensing, over-the-air synchronization, distributed processing, and AI in the loop.



\end{abstract}


\section{Introduction}
The International Telecommunication Union (ITU) is currently working on the next-generation standard for cellular communication called IMT-2030 (a.k.a. 6G) \cite{imt-2030}.
6G is envisioned to deliver significant advancements over existing 5G networks. While core functionalities like data rate, latency, and network capacity will see enhancements, the true distinction lies in the introduction of novel capabilities\cite{itu-r2160}. These include:
\begin{itemize}
    \item Ubiquitous intelligence: Seamless integration of artificial intelligence (AI) into communication networks, enabling intelligent automation and resource management.
    \item Focus on sustainability: Environmentally conscious network design and operation to minimize the environmental impact of ICT infrastructure.
    \item Universal connectivity: Bridging the digital divide by ensuring broader and more equitable access to communication services, particularly in underserved regions.
\end{itemize}


Given the focus on AI and integrated sensing and communication, it is essential to build platforms that allow rapid prototyping and implementation of new algorithms and the real-time collection of labeled datasets.
In our vision, these platforms should be as open and accessible as possible and provide a high degree of integration and interoperability. These goals align perfectly with European projects such as 6G-BRICKs\cite{6g-bricks} and SUNRISE-6G\cite{sunrise-6g}.

The scientific literature finds many works about designing and implementing experimental infrastructure for wireless communication research.
To this end, the authors of \cite{road-6g} make an excellent overview of the vision towards 6G, the projects currently addressing it, and how test facilities are being remodeled or built anew to tackle the new research challenges.
In \cite{distribuited_testbed, exp-building,design-validate}, the authors thoroughly document the design of their Testbed, the key parameters that are considered when designing and building them, and how they are using them effectively to target their specific needs.
With our Testbed, we intend to build an agile test facility to collect and share RF and positional datasets with the broader research community following the models of \cite{rfdatafactory,deepsense6g, eurecom_testbed}.
In particular, we closely cooperate with Eurecom \cite{eurecom_testbed}, which is also part of the 6G-BRICKs consortium.

    


The O-RAN Alliance \cite{oran-alliance} is an industry-driven effort promoting the transformation of Radio Access Networks (RAN) towards open and intelligent architectures. 
The initiative aims to deliver a more cost-effective, flexible, and high-performing RAN ecosystem for the mobile telecommunications industry. 
Their approach emphasizes disaggregation, allowing multi-vendor deployments and fostering innovation through open interfaces and software-defined components. In particular, O-RAN follows a service-oriented architecture, where many elements can be deployed as cloud services rather than local programs. This means, for example, that in a base station deployment, the Radio Unit (O-RU) can be from one vendor, and the Distributed Unit (O-DU) can come from another, potentially even running on different hardware and accessible remotely. The initiative aims to deliver a more cost-effective, flexible, and high-performing RAN ecosystem for the mobile telecommunications industry.

The success of our Massive MIMO Testbed, described in Section~\ref{subsec:mamimo-tb}, was the real-time labeled logging of CSI and positions for settings with a lot of freedom in antenna deployments (from URA to ULA to cell-free). To migrate to 6G, several functional additions to the Testbed are needed. Advanced sensing labeling is necessary for ISAC, ideally achieved using a motion capture system. Future standards will also have higher bandwidths (400 MHz or more) and higher frequencies (FR3 and FR2 for sensing).

These considerations and other requirements pushed us towards the design of a new Testbed, as described in Section~\ref{subsec:newtb}.

\subsection{Design objective for the new Testbed}
\label{subsec:newtb}
The design of the 6G Testbed at KU Leuven prioritizes several objectives to foster a collaborative and open research environment.
First, the Testbed leverages open-source software, ensuring transparency and enabling modifications by the research community.
Conversely, one of the goals of our research group is to also give back to the broader community by contributing to open-source initiatives such as Open Air Interface \cite{OAI-5gnr} and providing open and well-labeled datasets of physical layer measurements.
Second, remote access and participation in the upcoming European Federation of Testbeds promote broad participation and geographically distributed research efforts.
Third, interoperability with third-party components is a core objective, creating a modular ecosystem for experimentation and enabling collaboration with industrial partners.
Finally, the Testbed is designed for ease of use by employing containerized experiments and simple APIs to hide the complexity of the underlying hardware when it is not needed.





\subsection{Main contribution and structure of the paper}
The main contribution of this work is the design and development of an open experimentation facility, based on open-source software, with the aim of providing open public datasets.
This paper is divided in two parts: Section~\ref{sec:testbed_architecture} describes the design of the KU Leuven 6G Testbed, its capabilities, and the integration of the new test infrastructure with the preexisting Massive MIMO Testbed;
Section~\ref{sec:research-apps} describes the current research being performed in our group and how it maps to the testbed functionality.

\section{Architecture of the Testbed}
\label{sec:testbed_architecture}
\begin{figure}[t]
    \centering
    \includegraphics[width=\linewidth]{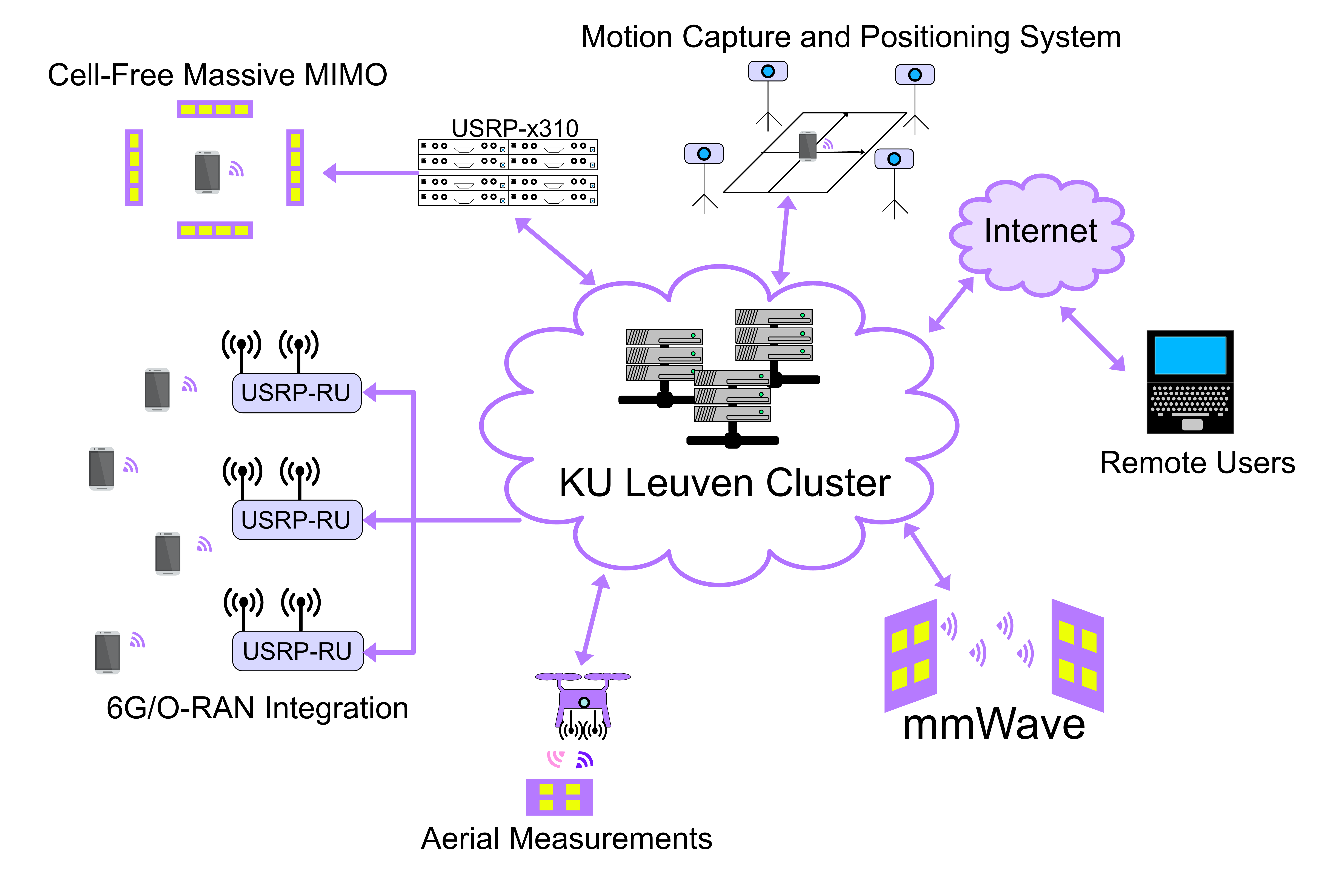}
    \caption{
    The Testbed revolves around a central cluster of machines and USRP software-defined radios. The cluster comprises multiple machines: a gateway for remote access and container deployment, a controller to drive the USRPs and perform real-time signal processing, and a compute server for long-running machine learning tasks.}
    \label{fig:testbed_architecture}
\end{figure}
The elements of the new test bed are shown in Figure~\ref{fig:testbed_architecture}.
It combines advanced hardware and FPGA-accelerated radios (USRP x310 and x410) with a data-logging and labeling compute server. The compute server, an HPE Proliant DL380, is closely integrated with advanced radio units and can quickly run communication functionalities (e.g., O-DU and RIC containers.)
The O-RU processing runs on the USRPs. 
The details about the USRP x410, which is the latest addition, are provided in Section~\ref{subsec:hardware}.
The USRPs can be synchronized using a clock generator and distributor called Octoclock. The Octoclock can use an internal or external reference, such as an atomic clock. Octoclock also includes a GPS receiver to take GPS signals for reference. 
Other notable components are a set of sixty-four modular antennas, a prototype mmWave front-end by Pharrowtech, and a DJI Inspire 2 drone equipped with a USRP E320.
Modular antennas are essential for experimenting with different array configurations.
The mmWave front-end is mainly used to research ISAC. 
Finally, the DJI drone takes aerial measurements outdoors and is used to research air-to-ground and air-to-vehicle communication.

\subsection{Compute Cluster for the Physical Layer}
The central component of the new Testbed is the KU Leuven cluster, comprising a Testbed controller and a gateway. The gateway runs the cloud infrastructure and security software, making the Testbed accessible online.

The Testbed controller connects to the radio units, an HP Proliant server with two Xeon Gold 6430 processors, 1TB of DDR5-4800 ECC RAM, and an Nvidia A100 AI accelerator. Each CPU has 32 cores and can run 64 threads simultaneously with a 60MB cache and Hyper-Threading. This enables efficient processing of O-RAN Distributed Unit (O-DU) tasks by splitting physical layer processing into multiple pipelined blocks.


\begin{figure}
    \centering
    \includegraphics[width=\linewidth]{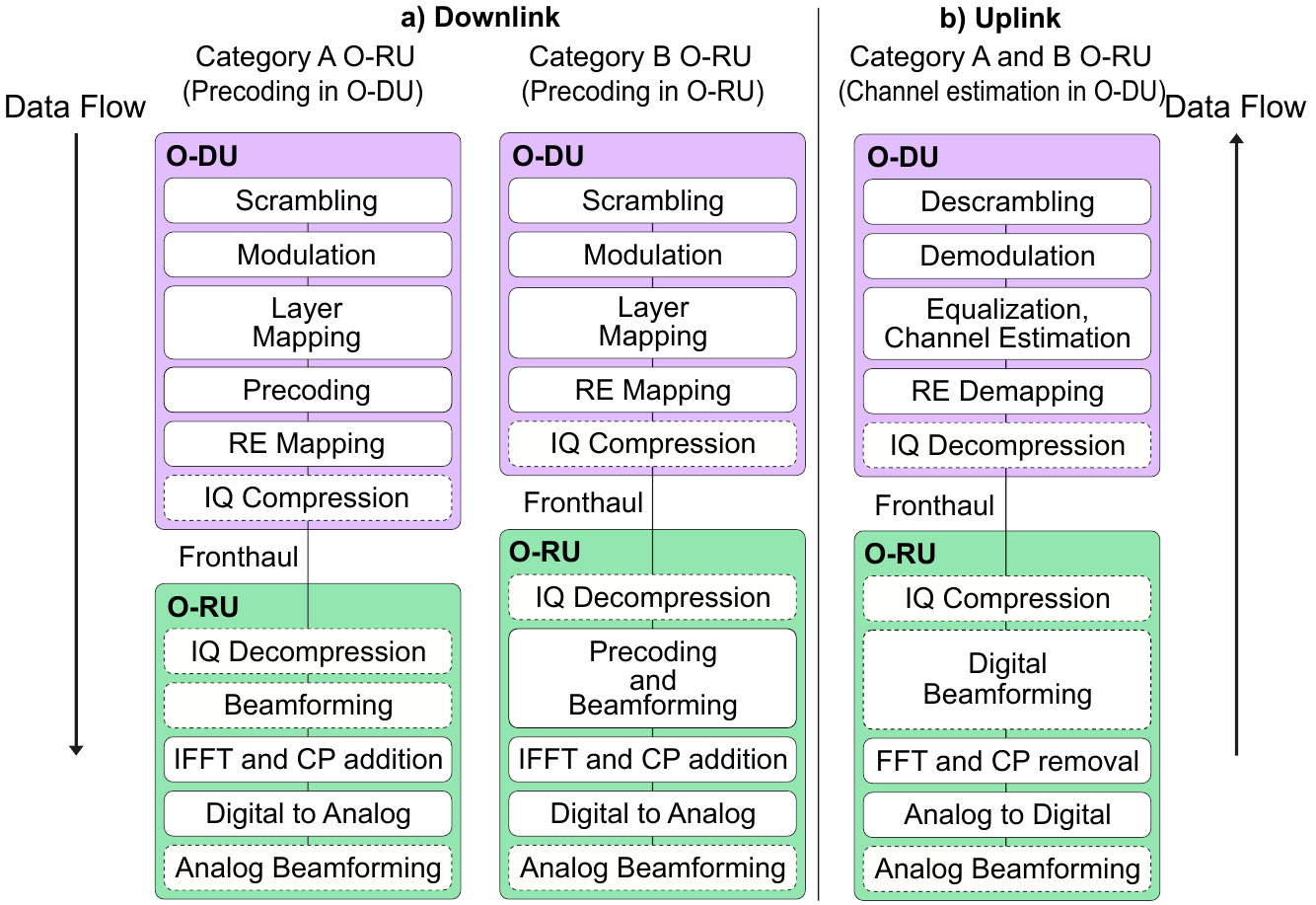}
    \caption{Different types of processing pipelines for O-RAN splits 7.2a and b. Depending on the vendors, stages can be distributed differently between O-RU and O-DU. \cite{vidaMaga}}
    \label{fig_O-RU_types}
\end{figure}

In Figure \ref{fig_O-RU_types}, we can see the different stages of the processing pipeline. 
We assume that part of the processing, mainly the Fast Fourier Transform of incoming and outgoing signals, can be efficiently performed in the O-RU. This combination is identified in O-RAN as split 7.
However, having many cores and fast RAM allows us to run the whole physical layer in software, a situation identified in O-RAN as split 8.
In the latter case, the various pipeline stages can be implemented as separate processes running in parallel on the many available cores.

Another essential characteristic of the Xeon Gold 6430 is the support in each core for the AVX512 instruction set. AVX512 contains instructions for working on eight 64-bit or sixteen 32-bit floating point numbers in parallel (SIMD instructions). This comes in handy when performing matrix operations, such as pre-coding, beam forming, and channel equalization, especially considering that antennas on the USRPs are in groups of four. 

Cache size is also vital for computation speed. Most physical layer operations can be arranged as vector operations over large arrays tightly packed in memory. This allows the processor's prefetch unit to load the data efficiently in the cache. However, to take advantage of the large cache, the programmer needs to adopt a "cache friendly" style of programming such as data-oriented design, where structs of arrays are adopted, instead of arrays of structs \cite{acton, soa-aos}.

Finally, the A100 Tensor Core accelerator allows us to perform training of machine learning models on the data acquired by the radios, inference on pre-trained models, perform research on ML in the loop of the physical layer, and finally to do massively parallel computations such as LDPC codes \cite{cuda-ldpc}. 

The controller is connected to the USRP-based O-RU via a 100G Ethernet link. 
The need for 1TB of RAM arises because disk read/write operations, even with fast SSDs, are much slower than RAM. At total bandwidth (400MS/s and 64 bits per sample), 3.2GB of data is generated per second. With half of the RAM dedicated to storing signals, the system can hold about 2.5 minutes of continuous data.

\subsection{Advanced Physical Layer Hardware}
\label{subsec:hardware}
\textbf{Distributed high-bandwidth PHY}: 
The hardware used to implement the O-RU is USRP x410. The x410 developed by NI Ettus Research is a high-performance software-defined radio. It utilizes a two-stage superheterodyne architecture, effectively converting and filtering frequencies from 1~MHz to 7.2~GHz, tunable up to 8~GHz. It features four independent transmit (TX) and receive (RX) channels, each capable of handling 400 MHz of bandwidth, enabling simultaneous processing of a wide frequency band. Each USRP X410 interfaces with the controller using two QSFP28 interfaces capable of 100 GbE for fast data transfer. The X410 is built on the Xilinx Zynq Ultrascale+ RFSoC that contains an ARM quad-core Cortex A53 CPU, an UltraScale+ FPGA, and an ARM Cortex-R5F real-time processor. The programmable logic (FPGA of the SoC) handles all sampling data, high-speed network connections, and other high-speed utilities such as custom RF Network-on-chip (RFNoC) blocks. 
RFNoC is an open-source framework for developing high-throughput DSP in the FPGA for USRP devices. It provides all the necessary infrastructure to integrate signal-processing IP blocks into the FPGA logic and interface with them via software. RFNoC splits processing into blocks (RFNoC blocks), and data is passed between the blocks. The RFNoC blocks wrap the custom IP and provide an interface to the rest of the infrastructure.
The processing system of the SoC runs a custom-built OpenEmbedded-based Linux operating system. Compared to the USRP X310, the USRP x410 offers a broader frequency range, greater instantaneous bandwidth, more independent channels, and improved processing capabilities due to the enhanced FPGA present in the RFSoC. 

\textbf{mmWave PHY}: 
The mmWave Testbed comprises of 
two main components: 
Baseband and RF front-end modules. 
The baseband module is built on the Zynq UltraScale+ RFSoC~\cite{rfsoc}, integrating multiple AD/DA converters capable of giga sample rates, meeting high throughput requirements in mmWave communication. The RF front-end modules include four pairs of Tx/Rx units: two pairs of EVK06002 development kits from Sivers~\cite{sivers} and two pairs of PTM1060 from Pharrowtech~\cite{pht}, with both supporting MIMO application and operating in the unlicensed 60~GHz band.  
The Sivers front-end includes two rectangular phased arrays for transmission and reception, each with $4\times16$ elements. Meanwhile, the Pharrowtech front-end comprises an $8\times8$ element-sized array. Additionally, fast beam switching capability is available via GPIO interfaces.

\subsection{KU Leuven distributed Massive MIMO Testbed}
\label{subsec:mamimo-tb}
The Massive MIMO Testbed at KU Leuven uses the NI USRP-2942R Software Defined Radio (SDR) platform and the Labview Communication Framework by National Instruments. The Base Station (BS) features a modular 64-antenna array, configurable in various geometrical setups like linear, rectangular, and distributed arrays. It includes 32 SDRs, each with two RF chains, functioning as remote radio heads to handle IFFT, digital-to-analog conversion, upconversion for downlink transmission, and inverse operations for uplink reception. Centralized signal processing is performed by a Baseband Processing Unit (BBU) using techniques such as maximum ratio, minimum-mean-squared error, or zero-forcing.

On the User Equipment (UE) side, the Testbed supports up to 12 UEs, each driven by a USRP-2942R connected to a host computer for signal processing. Synchronization with the BS SDRs can be achieved via over-the-air or coax cables for precise channel estimation. Automatic channel state information capture is enabled through a custom API, allowing a high degree of automation when performing measurement campaigns.

The MIMO Application Framework by NI, implemented in LabVIEW, supports BS-UE communication using Time Division Duplexing (TDD) based on the LTE frame structure. Integrated components include CNC XY-positioners for accurate UE antenna positioning, 
and a compact USRP E320 mounted on a DJI Inspire 2 UAV for aerial user experiments. The E320 includes internal processing, external amplifiers for improved SNR, and synchronization via over-the-air methods and GPS-locked oscillators.

The primary use of this Testbed is to collect real-time channel estimates for UL Massive MIMO and UL Distributed MIMO channels, facilitating diverse research areas such as signal processing, resource allocation, channel modeling, localization, and sensing, including human blockage prediction. The Testbed's versatility and the collected data have significantly contributed to advancing wireless communication research, as detailed in numerous research publications. 
The collection of channel estimates~\cite{mimoMagazine, sibrenDataset, achielDrone} enables highly diverse research such as signal processing~\cite{andreaInter, vidaMaga}, resource allocation~\cite{cmGrouping,achiel_mmw}, channel modeling~\cite{sanderDrone, zachDrone}, localization and tracking~\cite{chenglongTracking, chenglongLocalization}, sensing~\cite{adhamNearField}, including human blockage prediction~\cite{rizqi_guardbeam,bram_blockage22,bram_blockage23}.

\subsection{Data labeling}  
\textbf{Motion capture system:}
The system is built on the Qualisys Miqus motion capture camera, specifically the Miqus M3 model \cite{miqus}. This is a marker-based multi-camera motion capture system that provides high-precision tracking capabilities. One of its key features is offering diverse real-time API connections, supporting integration with various programming environments such as Python and MATLAB. This flexibility allows for seamless incorporation of motion capture data into radio development workflows.

\textbf{XY-positioning}: The Testbed is integrated with four CNC XY-positioners, which allows for highly accurate positioning of UE's antennas or other objects. Using the positioners, a fully automated measurement setup is built where the UE's antenna is moved over the region of interest. At the same time, channel state information is recorded, creating dense data sets of xy-location labeled channel state information that can be used for designing localization algorithms \cite{mimoMagazine}.

The motion capture system and the XY-positioner are linked to label the RF measurements accurately.

\subsection{Software}
The Testbed is built to support different software,  enabling different experiments. 
For our research, we primarily collect data about the channel state and then test physical layer algorithms on the collected data. 
\begin{figure}
    \centering
    \includegraphics[width=0.8\linewidth]{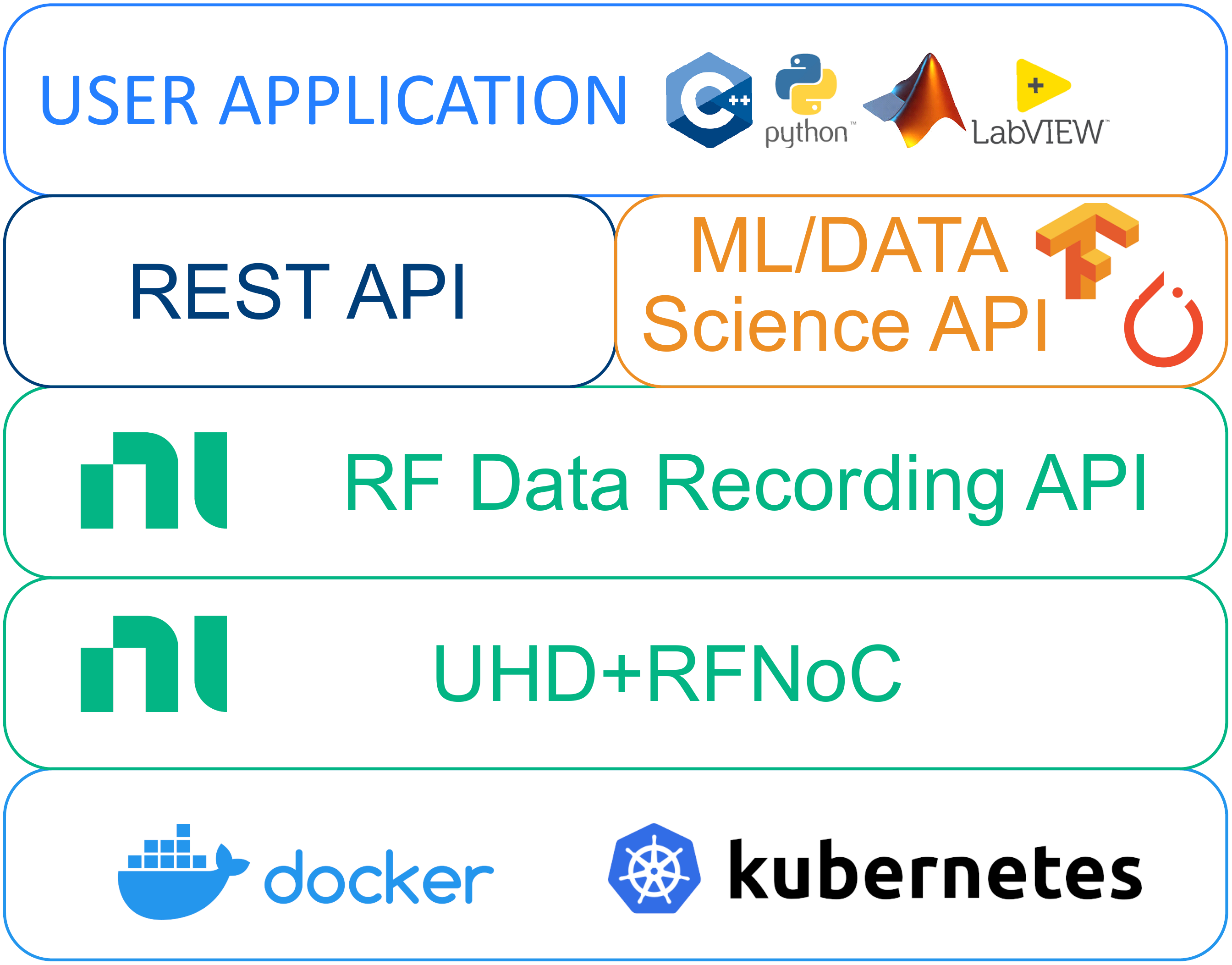}
    \caption{Basic software stack in use by the KU Leuven Cluster.}
    \label{fig:software-stack}
\end{figure}

Figure~\ref{fig:software-stack} represents the software stack we currently use.
In particular, National Instruments provides the open-source RF data recording API\cite{rf-datarecording,rf-datarecording-github}, a Python library to perform automatic measurements with the USRPs. 
The API is built around USRP models x310 and x410, but it can work with any other model supporting a recent version of the USRP Hardware Driver (UHD). 
The data collection API can be used to collect data that can be analyzed using frameworks such as PyTorch or TensorFlow.
User applications can access the hardware and the data collection API using our custom-built REST API.
The REST API is crucial in decoupling user applications from the underlying low-level libraries.
Finally, each experiment can be deployed as a Docker container, utilizing either the basic stack or a custom solution by adding and removing functionalities as necessitated by the specific requirements of the situation.


\section{Research Applications}
\label{sec:research-apps}


\subsection{Over-The-Air synchronization}
A significant challenge in distributed communication systems is synchronizing all antennas in phase and time for precise beamforming. Each device has its local oscillator, causing time and phase shifts due to imperfections and noise. Over-the-air (OTA) synchronization ensures accurate timing across wireless fronthauls, eliminating the need for physical connections, which become unscalable with more Access Points (APs). This method provides a flexible, cost-effective solution for deploying wireless APs with accuracy comparable to wired solutions.

Using our Testbed, we aim to implement and evaluate OTA synchronization mechanisms across distributed USRP devices, addressing challenges like signal interference and network latency. These mechanisms can use dedicated frames (such as IEEE 1588/PTP) or designated signals.

The ideal solution \cite{5gsync} involves wirelessly distributed Radio Units (RUs) and some wired Distributed Units (DUs) and RUs. It combines methods like GNSS and OTA synchronization at base stations, absolute time and frequency distribution from the transport network, and redundancy support. The research will progressively unlink radio devices, comparing each situation until achieving a fully wireless fronthaul. We aim to find the optimal OTA synchronization architecture for a wireless fronthaul cell-free network while maintaining Open RAN compliance.

\subsection{Integrated sensing and communication} 




Integrated Sensing and Communication (ISAC) is a key feature anticipated in 6G networks, enabling the simultaneous provision of sensing services alongside user communication ~\cite{Fang2023_isac}. Our Testbed infrastructure supports ISAC research across multiple frequency bands and sensing modalities, leveraging our mmWave and Massive MIMO setups.
\begin{itemize}
  \item \textbf{Multi-band sensing:} 
 Multi-band sensing combines sub-6 GHz's wide coverage and penetration with mmWave's high-resolution sensing, providing robust and precise sensing across diverse environments.
  \item \textbf{Multi-modal sensing:} 
  Our Testbed supports advanced multi-modal sensing research by integrating cellular signals from massive MIMO Testbeds, UAV aerial signals, mmWave high-frequency signals, high-resolution motion capture visuals, and precise CNC XY-positioner positioning, enabling the development of comprehensive multi-modal sensing algorithms.
  \item \textbf{Customizable ISAC waveforms evaluation:} 
  Our Testbed's flexibility, featuring the Zynq UltraScale+ RFSoC in the mmWave setup and the software-defined Massive MIMO Testbed, enables the design and testing of novel ISAC waveforms. The mmWave phased-array frontend supports customizable beamforming, allowing researchers to explore waveforms that balance communication and sensing needs.

\end{itemize}

By providing this comprehensive ISAC research platform, our Testbed enables the exploration of sensing applications ranging from human presence detection and positioning to activity recognition and vital signs monitoring ~\cite{HaoqiuLocl,HaoqiuVital} across multiple frequency bands and sensing modalities. This multi-faceted approach aligns with the anticipated requirements of future 6G networks, where ISAC is expected to play a crucial role in enabling new services and applications.

\subsection{Distributed processing}

Distributed processing has been the focus of many papers on telecommunication networks in which multiple-input, multiple-output (MIMO) technology is used on the transmitter and receiver side. The reason is to offload the substantial computational load from a central processing unit and distribute the processing capabilities among the APs \cite{bjornson_making_cell_free}. Distributed MIMO networks can be realized using different fronthaul architectures such as daisy chains or star topology. Daisy chain implementation of a MIMO network, also called radio stripe architecture \cite{Internado_radio_stripe}, is an ideal architecture in places such as sports arenas and historical touristic places. Besides that, the fronthaul traffic to a single element is reduced as opposed to the star architecture, where a central element is congested with a tremendous amount of traffic. Furthermore, having a daisy chain architecture, optimal algorithms for uplink signal estimation can be deployed among APs with the least amount of information exchange \cite{shaikh_MMS_dis, ranjbar_recursive}.
    
Despite being rich in theoretical aspects, having a realistic MIMO Testbed where distributed processing is done in practice is still missing. 
The Testbed's architecture allows the radio units to be connected in different configurations (e.g., daisy chain and multi-branch tree). Various interconnection schemas are essential to study the impact of different non-idealities, such as constrained fronthaul links \cite{chiotis_rs, ranjbar_eucnc_2024} or limited working memory \cite{ranjbar_limited_memory_tcom} at the APs, on the quality of service of the users. 

\subsection{AI in the loop} 
A promising direction in communications research involves enhancing and eventually replacing conventional functionality with AI-based methods\cite{hoydis_AInative}. This evolution leverages artificial neural networks for tasks such as synchronization, demodulation, channel estimation, channel coding, and data compression, driven by improvements in spectral efficiency, reliability, and energy efficiency\cite{trainable_comms, csifeedback, joint_detection_decoding}. Beyond the physical layer, AI optimizes the overall network using RIC controllers, which centralize network management and enhance performance through optimized resource allocation, scheduling, load balancing, and handovers\cite{xapps}.

The proposed testbed infrastructure supports realistic dataset collection and offline AI training, as well as deploying pre-trained ML models across the protocol stack. Depending on the application's compute, latency, and power requirements, these models can be deployed on a commercial GPU in the O-DU software or as a custom RFNoC block in the O-RU low PHY. While separate AI-based functional blocks have been validated in over-the-air experiments\cite{trainable_comms, joint_detection_decoding}, their integration into a realistic Open Air Interface implementation poses challenges such as automated ML model deployment, lifecycle management, and neural architecture compression.

\section{Conclusions and Future Work}
This work describes our effort in designing and building a test facility for future 6G physical layer research. 
The requirements, architecture, and selected components are described in detail, together with the rationale for each technical choice. 
Finally, we provide a detailed discussion on current and potential research applications that can be explored using the Testbed. 
In the future, we aim to expand the testbed with new functionality, such as different radio front-ends, different antenna arrays, and the hardware implementation of new signal processing blocks.

\bibliographystyle{ieeetr}
\bibliography{refs}

\end{document}